DIAGNOSTICO SITUACIONAL DE LOS DOCENTES DE PRIMARIA EN FORMACIÓN SOBRE ALGUNOS FENOMENOS ASTRONÓMICOS

Alejandro Gangui, María C. Iglesias, Cynthia P. Quinteros
IAFE – CONICET  y  CEFIEC – FCEyN – UBA

relat@iafe.uba.ar

El estado de situación de la enseñanza-aprendizaje en temas de astronomía es un campo de investigación poco explorado en nuestro país. Esto es así a pesar de reconocerse en la astronomía una de las disciplinas integradoras por excelencia, punto de encuentro de los avances en física, geología y química, entre varias otras áreas del saber. En nuestro proyecto nos proponemos contribuir al diagnóstico situacional de los docentes de escuela primaria en formación, con el propósito de desarrollar herramientas didácticas que contribuyan a mejorar su educación formal. Trabajamos con un cuestionario escrito y abierto –aunque también se incluyen algunas preguntas cerradas– diseñado con el fin de evidenciar los modelos de explicación más frecuentes para un grupo selecto pero representativo de nociones básicas de astronomía. Estudios internacionales recientes mostraron que no sólo los alumnos de la escuela primaria, sino también sus futuros docentes, llegan al aula de ciencias con modelos previos (aunque no científicos) del universo que los rodea. Estas ideas, especialmente en temas de astronomía, se revelan robustas y en general obstaculizan el proceso de enseñanza-aprendizaje. Si no son tenidas en cuenta –si no se trabaja a partir de ellas para construir nuevos conocimientos– el aprendizaje de conceptos científicos se torna lábil y no significativo. En esta nota se discuten varios de los temas de astronomía que se analizan: gravitación y verticalidad, estaciones del año, ciclo día-noche, movimiento y fases de la Luna, eclipses, sin olvidar aspectos culturales y terminología corriente basada en la astronomía que se da en la vida cotidiana.

Área temática: Didáctica de las Ciencias de la Naturaleza.

Introducción

La importancia otorgada por los diseños curriculares a temas relacionados con los fenómenos astronómicos se evidencia en muchas de las jurisdicciones de nuestro país. Tomaremos como ejemplo a la Ciudad Autónoma de Buenos Aires por tratarse de la zona de interés de quienes realizan esta investigación. El Bloque *La Tierra y el Universo* nuclea los contenidos referidos, entre otros, a ciertos fenómenos astronómicos. La organización de contenidos deja entrever claramente la progresión propuesta a lo largo de ambos ciclos, como así también sus alcances. Los alumnos pasan del reconocimiento de regularidades y cambios (duración de días y noches en distintas estaciones, cambios en aspectos de la Luna, etc.) en el primer ciclo, para comenzar a formarse, a partir de 5º Grado, una imagen más estructurada de la Tierra y del Universo, dejando para el último año de la primaria las relaciones más complejas relativas al Sistema Solar. Esto ha llevado a cuestionarnos acerca de la formación –no específica– que presentan los futuros docentes de enseñanza primaria, particularmente en esta jurisdicción, y por lo tanto sobre la forma en que los diversos contenidos de astronomía son abordados dentro de aula. En un análisis posterior de los lineamientos curriculares para la formación docente, encontramos algunos de los temas que presentan muy arraigadas ideas previas, como ser la explicación de las estaciones del año, y donde las clásicas imágenes de elipses muy elongadas para las trayectorias de los planetas (como la Tierra) pueden inducir a los futuros docentes a aceptar, sin mucho cuestionamiento, la teoría del alejamiento para explicar las diferentes temperaturas o duraciones del día en una u otra estación del año.

Los modelos que construye la ciencia para explicar la realidad parten de las representaciones individuales de los científicos. De igual modo, los docentes de primaria en formación llegan al aula con modelos pre-construidos y consistentes del universo que observan. Es sabido que existe una gran variedad de temas de astronomía en los cuales los futuros docentes de la escuela primaria presentan ideas previas. Esta constatación surge de diversos estudios realizados en varios países (ver, por ejemplo, Callison y Wright, 1993; Camino, 1995; Atwood y Atwood, 1995; Atwood y Atwood, 1996; Parker y Heywood, 1998; Vega-Navarro, 2001; Trundle *et al.*, 2002; Trumper, 2003; Martínez-Sebastià, 2004; Frede, 2006). En algunas de estas investigaciones se señala que la recurrencia de las ideas previas en relación a estos temas se prolonga más allá de la escuela secundaria, afectando también a estudiantes de magisterio y profesores de enseñanza primaria. En nuestro proyecto presentamos una serie de interrogantes básicos que, en su conjunto, llevan a reflexionar acerca del estado de situación de la enseñanza-aprendizaje en astronomía en el ámbito de la educación formal. Se plantea, a su vez, la necesidad de dar continuidad, en nuestro país, a investigaciones en el área de la didáctica de la astronomía.

Descripción de la investigación

En este proyecto trabajamos con los docentes de escuela primaria en formación. Creemos que es la población más relevante para comenzar nuestro estudio de diagnóstico y posterior desarrollo de herramientas para un aprendizaje significativo. El objetivo principal de esta investigación es contribuir al diagnostico situacional de docentes en formación en relación a temas de astronomía (Gangui, 2007; Iglesias *et al.*, 2008). Nuestros objetivos particulares son: 1) indagar el estado de conocimiento de la población docente del nivel primario en formación en relación a temas de astronomía, 2) analizar las representaciones de los futuros docentes y que actúan como obstáculos para el aprendizaje de contenidos relacionados con el área, 3) desarrollar herramientas didácticas innovadoras con participación conjunta entre formadores de docentes e investigadores. En este artículo describimos algunos de los elementos diseñados para llevar a cabo la primera fase de este estudio, y discutimos algunos resultados preliminares.

Instrumento

Describiremos aquí brevemente el cuestionario que usamos para indagar algunas ideas previas de los docentes del nivel primario en formación. Este consta de 12 preguntas que cubren unas ocho temáticas diferentes en astronomía básica (ver *Tablas* 1 a 4 en las páginas siguientes). Las primeras dos preguntas indagan sobre las ideas de verticalidad y de la gravitación como una fuerza de atracción hacia el centro de la Tierra (Nussbaum, 1979; Sneider y Ohadi, 1998). La referencia a una historieta en la primera pregunta busca distender y dar una idea de familiaridad al interrogado. En particular, el comentario sobre los habitantes con la cabeza hacia abajo sin duda expresa una idea común que, sea en broma o en serio, muchos han oído alguna vez. En la instancia de explicación y/o justificación de la respuesta dada, el interrogado se plantea la situación nuevamente y muestra (ya sea con un breve texto o con un dibujo) su concepto natural de verticalidad. En la segunda pregunta, se trata de hacer que el interrogado se replantee la situación de la acción de la gravitación en la Tierra. A partir de su respuesta el investigador puede verificar si su interpretación de la pregunta 1 era casual o si, por el contrario, se trata de una estructura conceptual estable.

> 1) En una historieta de Malfalda ella se encuentra preocupada por nuestra ubicación en el globo terráqueo. Cuando descubre dónde estamos, se desespera al pensar que vivimos cabeza abajo. ¿Es cierto que los habitantes del hemisferio sur estamos con la cabeza hacia abajo? Explique.
>
> 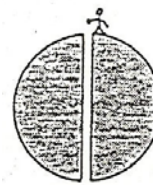
>
> 2) Suponga que se construye un túnel que atraviesa la Tierra diametralmente, ¿cuál sería el movimiento de una bolita que se deja caer en una de las bocas de dicho túnel? ¿Por qué?

*Tabla 1: preguntas que buscan indagar ideas sobre verticalidad y gravedad.*

Luego se pasa a un grupo de tres preguntas que permiten indagar las ideas sobre conceptos astronómicos simples propiamente dichos, como ser el ciclo día-noche, el movimiento propio de nuestro planeta, las diferentes estaciones del año, la órbita que la Tierra describe alrededor del Sol, el grado de inclinación del eje del planeta con respecto a su plano de movimiento (la eclíptica), todos ellos, de una manera u otra, muchas veces puestos en relación (en dependencia) y a veces confundidos a la hora de brindar una explicación científica (válida) a cómo se producen los fenómenos astronómicos que observamos, ya sea durante el transcurso del día o al completarse un año (Parker y Heywood, 1998). Nuevamente, la pregunta 5 pretende analizar el grado de consistencia de la respuesta formulada en la pregunta 3.

> 3) ¿Por qué hay diferentes estaciones en el año?
>
> 4) Explique el ciclo día-noche, es decir, por qué hay días y noches en la Tierra.
>
> 5) ¿Por qué hace más calor en verano que en invierno en el hemisferio sur?

*Tabla 2: preguntas que buscan indagar ideas sobre conceptos astronómicos simples.*

Las preguntas 6 a 8 están dedicadas a indagar sobre las ideas que guardan los futuros docentes sobre los movimientos que posee un cuerpo astronómico muy familiar como lo es la Luna. La mención explícita a una única cara visible de la Luna (pregunta 6) permite indagar si el encuestado relaciona adecuadamente la combinación de movimientos necesarios para explicar el movimiento real, cuasi mensual, del astro. La pregunta 7 introduce en el razonamiento también al Sol, al movimiento de este sistema de tres cuerpos y a cómo efectivamente se relacionan fenómenos bien conocidos –los eclipses– con la disposición espacial (y temporal también, por supuesto) de esos cuerpos. Esta pregunta guarda relación también con la pregunta 10, sobre las fases lunares, pues se ha visto en diversos estudios que estas fases son muchas veces interpretadas como el resultado de la interposición de la Tierra en el camino de la luz solar que, de otra manera, iluminaría la superficie de la Luna (teoría del eclipse; ver, por ejemplo, Camino, 1995). La pregunta 8 fue *ex profeso* separada de la 6 para que la visión de aquella no incidiese en la posible respuesta de esta última. En efecto, ambas preguntas están muy relacionadas, aunque la 6 tiene algunas reminiscencias culturales mientras que la 8 puede resultar un poco más árida y complicada. Sabemos, sin embargo, que la adecuada respuesta a la 8 implica un acabado conocimiento de lo que se quiere indagar en la 6. Se busca también que, al llegar a la pregunta 8, el encuestado reconsidere su respuesta de la 6 en busca de reafirmación de lo que piensa. Esto sirve a la investigación, pues permitirá detectar mejor la verdadera concepción del futuro docente en lo que hace a la relación Tierra-Luna y al movimiento coordinado (subordinado) de esta última.

La pregunta 9 busca explorar las ideas que los docentes en formación tienen sobre los verdaderos movimientos de la Tierra. Tiene conexión directa con las preguntas 3, 4 y 5, planteadas previamente.

6) ¿Cuál es la razón por la cuál una persona siempre observa la misma cara de la Luna?

7) En la madrugada del jueves 21 de febrero de este año hubo un eclipse de Luna, ¿qué condiciones deben cumplirse para que esto ocurra?

8) ¿Qué movimientos posee el cuerpo de la Luna? Descríbalos.

9) ¿En qué consisten los movimientos de rotación y traslación terrestres? Descríbalos.

10) El día 12 de mayo de este año podremos observar la Luna en su cuarto creciente.
    a) ¿Cómo se produce esta fase de la Luna?
    b) ¿Y una fase de Luna nueva?

*Tabla 3: preguntas que buscan explorar las explicaciones relativas a la Luna y los movimientos de la Tierra.*

Las últimas dos preguntas, la 11 y la 12, sirven para indagar aspectos culturales de la astronomía básica, ya sea en forma de definiciones como a qué se llama comúnmente una estrella polar (Frede, 2006) o qué es realmente una estrella fugaz (pregunta 11). La pregunta 12, al igual que lo hacía la pregunta 1 (con otro estilo, por supuesto), nuevamente recurre a un texto para indagar el conocimiento que el encuestado tiene sobre frases o dichos de uso corriente en el lenguaje común. El poema plantea / explicita una situación clara en el atardecer. Se busca testear cuánto y cuán bien la astronomía integra el discurso cotidiano de personas no necesariamente relacionadas con la disciplina, pero que en sus clases luego podrán emplear formas de expresión que involucren aspectos astronómicos (Gangui, 2008).

> 11) ¿Qué es una estrella fugaz?
>
> 12) En el poema "Una despedida" de Jorge Luís Borges podemos leer la siguiente frase: *"La noche había llegado con urgencia. Fuimos hasta la verja en esa gravedad de la sombra que ya el lucero alivia."* ¿A qué se refiere Borges con "el lucero"?

*Tabla 4: preguntas referidas a aspectos culturales de la astronomía básica.*

Discusión

Puesto que nuestro trabajo se encuentra en la primera fase de la investigación, realizamos una prueba piloto del cuestionario a fin de "ajustarlo" y asegurar su cabal comprensión. Y, por otro lado, disminuir la posible intervención del encuestador a la hora de llevarlo a cabo. Para esto, se escogió una Escuela Normal Superior de la Ciudad Autónoma de Buenos Aires. El grupo "piloto" estaba conformado por 16 alumnas pertenecientes a las carreras de los profesorados de enseñanza primaria e inicial (de una materia que conforma el bloque común a ambas carreras). El motivo de esta elección radica en que, por tratarse de una prueba piloto, sólo se buscaba analizar si las preguntas eran debidamente interpretadas, considerando los intereses y objetivos de los investigadores. Por otro lado, existe la dificultad de agrupar, en un mismo momento, a varios alumnos (las clases de materias específicas para el profesorado de enseñanza primaria suelen estar integradas por pocos alumnos).

Lo primero que merece la pena destacar, es que el cuestionario se resolvió en un tiempo considerablemente menor al estimado previamente. Se suponía una demora de aproximadamente dos minutos por pregunta, tiempo que implicaría tanto la lectura e interpretación de la pregunta como la elaboración de una respuesta. En un primer momento, se pensó que las preguntas podrían ser de fácil resolución para las participantes. Sin embargo, en un análisis posterior se encontró que algunas preguntas no fueron respondidas. En especial, se encontró coincidencia para las preguntas referidas a la Luna. Ninguna encuestada se "animó" a dar una explicación sin sentirse segura de su respuesta. Y así, se justificaron con no recordar el tema en cuestión. Y muchas de ellas, aseguraron –y hasta se mostraron sorprendidas de– nunca haber notado que la Luna nos mostrara siempre la misma cara.

Por otro lado, para la primera pregunta no se encontró, en las respuestas, referencia alguna a la noción investigada. Esto nos induce a pensar que quizás el recurso de la historieta finalmente llevaría a que los encuestados no la considerasen una pregunta válida, sino poco seria (por estar enmarcada en una historia cómica). Asimismo, se encontró unanimidad en la respuesta ("No es cierto que estamos cabeza abajo") lo cuál podría explicarse como una superación del modelo no esférico de la Tierra, que se supone presente en los niños de corta edad. Sin embargo, esto no significa que el conocimiento de la dirección del campo gravitatorio de la Tierra sea una estructura conceptual estable. De hecho, ninguna encuestada logró dar siquiera una breve explicación. Al analizar sus respuestas a la pregunta 2, evidenciamos en algunos casos, una imposición de un nuevo movimiento para la Tierra con el cuál justificar el movimiento de la bolita que se deja caer. Y esto, claramente, no se relaciona con el concepto antes mencionado. Por ejemplo, encontramos que "la bolita va y viene por el movimiento que realiza la Tierra". Otras respuestas, no obstante, asumen un movimiento recto de la bolita debido a la gravedad aunque no describen en detalle ese movimiento. Por lo tanto, no puede saberse qué suponen que le sucede a la bolita al entrar al túnel, si sigue su curso o, incluso, si sale del túnel, y en consecuencia no podemos conocer sus modelos de explicación.

Algunas alumnas, además, no realizaron una interpretación correcta de la situación propuesta negando, inclusive, la posibilidad de que un supuesto túnel pudiera realizarse para dejar caer la bolita. Debido a todo lo expuesto, consideramos pertinente reformular ambas preguntas a fin de facilitar su comprensión y poner en evidencia sus modelos explicativos.

Respecto de las preguntas dirigidas a explorar los modelos utilizados para explicar las estaciones del año (3 y 5) encontramos una causalidad inversa. Muchas alumnas, en lugar de proponer un modelo explicativo –causa– que justificara un determinado fenómeno –efecto–, hacían uso de otro fenómeno-efecto, en este caso en relación al clima, fenómeno que, como sabemos, posee una componente astronómica.

*Ejemplo de respuesta a la pregunta 3*: "Por los distintos climas".

*Ejemplo de respuesta a la pregunta 5*: "Debe ser por el tipo de clima de este hemisferio".

Algunas de las otras respuestas para esta serie de tres preguntas, como se esperaba, hacían uso de la teoría de alejamiento (Camino, 1995).

En el caso de las preguntas en relación al ciclo día-noche no se evidenciaron inconvenientes para su interpretación, y el modelo más frecuente se corresponde con el científico aceptado hoy en día. Sin embargo, algunas pocas alumnas apelaron a un modelo geocéntrico, no sólo para el ciclo día-noche sino también para las estaciones del año, aunque en este último caso lo adjudicamos a una mala (o poco clara) redacción. También consideramos pertinente notar que, para la pregunta 4, si bien encontramos respuestas que hacen uso del modelo de rotación de la Tierra, para algunas alumnas este hecho conduce a que el Sol se aprecia de día y la Luna –exclusivamente– de noche.

Para las preguntas 11 y 12, que indagaban sobre algunas definiciones astronómicas corrientes y terminología propia de la vida cotidiana, obtuvimos respuestas de lo más variadas: desde "estrellas muertas" que penetran la atmósfera terrestre, como explicación de estrellas fugaces (pregunta 11), hasta toda una variedad de objetos luminosos (astronómicos o no) para la respuesta de la pregunta 12. Justo es mencionar, sin embargo, que también hubo respuestas muy bien encaminadas para la pregunta 11.

Para finalizar, queremos destacar que si bien se trata de una prueba piloto y, por lo tanto, una muestra no significativa, el análisis realizado nos deja entrever que las ideas previas y otras dificultades de aprendizaje, están presentes en nuestros futuros docentes y, por lo tanto, merecen especial atención. No obstante, nuestras próximas indagaciones permitirán un análisis mucho más profundo y reflexivo.